\documentclass[aps,nofootinbib,floatfix,twocolumn]{revtex4-2}
\usepackage{amsmath,amssymb,color,graphicx,bm,hyperref,mathrsfs}
\usepackage{physics}
\usepackage{verbatim}

\definecolor{red}{rgb}{0.8,0,0}
\definecolor{RED}{rgb}{0.8,0,0}
\definecolor{violet}{rgb}{0.4,0,0.4}
\definecolor{green}{rgb}{0,0.5,0.0}
\definecolor{GREEN}{rgb}{0,0.5,0.0}
\definecolor{navy}{rgb}{0.0,0.0,0.6}
\definecolor{orange}{rgb}{0.8,0.2,0.0}
\definecolor{blue}{rgb}{0.3,0.0,0.8}

\begin{document}
\title{\textbf{Thermal scalar field stress tensor on a two dimensional black hole and its near horizon properties}}

\author{Saurav Samanta$^{a,}$}
\email{srvsmnt@gmail.com}

\author{Bibhas Ranjan Majhi$^{b,}$}
\email{bibhas.majhi@iitg.ac.in}


\affiliation{$^a$Department of Physics, Bajkul Milani Mahavidyalaya, Kismat Bajkul, Purba Medinipur 721655, India.\\
$^b$Department of Physics, Indian Institute of Technology Guwahati, Guwahati 781039, Assam, India.
}

\begin{abstract}
We calculate the thermal renormalized energy-momentum tensor components of a massless scalar field, leading to trace anomaly, on a $(1+1)$ dimensional static black hole spacetime. Using these, the energy density and flux, seen by both static and freely-falling observers, are evaluated. Interestingly for both these observers the aforementioned quantities in the thermal version of Unruh and Boulware states are finite at the horizon when the scalar field is in thermal equilibrium with the horizon temperature (given by the Hawking expression). Whereas in Hartle-Hawking thermal state both the observers see finite energy-density and flux at the horizon, irrespective of the value of field temperature. Particularly in the case of Schwarzschild spacetime a freely falling observer, starts with initial zero velocity, finds its initial critical position $r^c_i = (3/2)r_H$, where $r_H$ is the horizon radius for which energy-density vanishes.  
\end{abstract}
\maketitle	

\section{Introduction}
In absence of ``complete'' quantum theory of gravity, semiclassical analysis appears to be fruitful in illuminating features of quantum nature of gravity. In this approach the background spacetime, representing gravity, is taken to be classical while the matter living on it is quantized. Most intriguing outcomes are Hawking \cite{Hawking:1975vcx} and Unruh \cite{Unruh:1976db} effects. Quantum aspects of matter on a gravitational background are being highlighted either through detector response \cite{Birrell:1982ix,Takagi:1986kn,DeWitt:2003pm} or through renormalized version of matter energy-momentum tensor (EMT) \cite{Birrell:1982ix,DeWitt:2003pm}. The latter approach has been succeeded in extracting Hawking flux \cite{Christensen:1977jc,Robinson:2005pd,Iso:2006ut,Banerjee:2007qs,Banerjee:2007uc,Banerjee:2008sn,Majhi:2014hpa,Banerjee:2019tbr,Seenivasan:2023pjd}, incorporating corrections to fluid constitutive relations due to presence of gravitational anomalies \cite{Son:2009tf,Landsteiner:2011cp,Jensen:2012kj,Jain:2012rh,Banerjee:2013fqa,Majhi:2014eta,Majhi:2014hsa,Banerjee:2014ita} and it also features the nature of energy-density and flux of matter as seen by different observers (see, \cite{Eune:2014eka,Kim:2014cja,Smerlak:2013sga,Kim:2013caa,Singh:2013pxf,Chakraborty:2015nwa}). The studies with massless scalar field in $(1+1)$-dimensional Schwarzschild black hole spacetime revels that the energy-density at the horizon as seen by a freely-falling frame, depending upon the initial position of the frame and choice of vacuum, can be divergent as well as finite ({\it{e.g.}} see \cite{Eune:2014eka}). In Boulware vacuum, it is negative divergent, whereas in Unruh it can be negative divergent as well as finite. However, in Hartle-Hawking (Kruskal) vacuum this is always finite and changes sign around a critical initial position $3M$, where $M$ is the mass of the black hole.   

Most of these studies are done with zero-temperature quantum field theory. However, quantum fields can be in thermal state. The investigation with thermal state is motivated from the following facts.
\begin{itemize}
\item The environment with finite temperature is more natural than zero temperature. So, the background fields in general can be thermal. There can be always a thermal noise around a black hole. For instance, the accretion disk around a Stellar-mass black hole has temperature as high as $\sim 10^{9}$ Kelvin \cite{Yuan:2014gma}. This even increases as the accreting matter goes near the horizon. Therefore fields in this region are dominantly in thermal state. 
\item The quantum phenomenon, like the response of a detector in the thermal state appears to be non-trivial \cite{Costa:1994yx,Kolekar:2013xua,Kolekar:2013aka,Chowdhury:2019set,Barman:2021oum,Barman:2022utm} and modified reasonably compared to the same due to vacuum fluctuations. Moreover, entanglement phenomenon between the quantum states of detectors are considerably influenced by thermal state of fields \cite{Huang:2017yjt,Barman:2021bbw,Chowdhury:2021ieg}. Even the entanglement between two Unruh-deWitt detectors is reduced due to thermal fluctuations of fields. 
\end{itemize}
Consideration of the above facts directs us to investigate the renormalised EMT.
In this regime, as far as we are aware, rigorous investigations on renormalized version of EMT corresponding to thermal quantum fields has not been done. There are few aspects to investigate in the structures of the thermal EMT components. We like to find the energy-density and the flux as seen by two kind of observers -- one is static and another one is freely falling towards the horizon. It would be interesting to see how the temperature of the field influences the properties of these quantities near the horizon. Considering the vast importance of this subject (see e.g. \cite{Eune:2014eka,Kim:2014cja,Smerlak:2013sga,Kim:2013caa,Singh:2013pxf,Chakraborty:2015nwa}), the inclusion of effects due to background temperature provides us more realistic results.

In this investigation we consider a massless scalar field at inverse temperature $\beta$ on a $(1+1)$-spacetime dimensional static black hole. At the quantum level, the conformal symmetry or both diffeomorphism and conformal symmetries can break down, leading to trace anomaly or both diffeomorphim and trace anomalies. Here we take the case where only trace anomaly appears and diffeomorphism symmetry is kept intact. This is consistent with semi-classical version of Einstein's equations of motion $G_{AB} = 8\pi G <T_{AB}>$. The components of renormalized EMT in the presence of $\beta$ are being evaluated in different thermal states. We take those thermal states which at the zero temperature (e.g. $\beta\to\infty$) lead to well known Unruh, Boulware and Hartle-Hawking (or Kruskal) vacua. Therefore we name these thermal states as Unruh therml state, Boulware thermal state and Hartle-Hawking (or Kruskal) thermal state. 

Finding of regularized components of energy-momentum tensor and its properties do not complete the job. In general these components depend on the choice of coordinates and hence their behavior can be due to pathological nature of the chosen coordinate system. However any scalar quantity, made out of these components, provides more physical insight in a coordinate invariant way. The two scalar quantities we introduce here are energy-density and flux as measured by an observer whose proper velocity is given by $u^A$. Similar quantities have been investigated earlier in \cite{Eune:2014eka,Kim:2014cja,Smerlak:2013sga,Kim:2013caa,Singh:2013pxf,Chakraborty:2015nwa} for zero temperature field theory.
Using the obtained components of stress-tensor, we find the energy-density and the flux, two locally defined observables \cite{PhysRevD.53.1988,Ford:1993bw,PhysRevD.51.4277}, as measured by a static and a freely-falling observers. Contrary to zero temperature field theory, here at the horizon of the black hole, these quantities can have regular behaviour if the matter is in thermal equilibrium with the horizon, i.e. the value of $\beta$ equals to inverse Hawking temperature. Particularly with respect to these two observers, the Unruh and Boulware thermal states lead to finite energy-density and vanishing flux at the horizon when such thermal equilibrium is present. On the other hand the Hartle-Hawking thermal state provides finiteness of these quantities which are independent of the value of $\beta$. In the static observer's frame, the energy-densities at the horizon of Schwarzschild black hole in all the thermal states are negative. On the other hand, the freely falling observer case shows a distinct feature depending upon its initial position. While the flux at the horizon is always positive, the energy-density can be both positive and negative around a critical value $r^c_i=(3/2)r_H$ of the initial position where its horizon value vanishes. When the initial position is larger than critical value, {\it{i.e.}} $r_i>r^c_i$ the energy-density at the horizon is positive and for other side ({\it{i.e.}} $r_i<r^c_i$) it is negative.  

The condition that matter is in thermal equilibrium with the horizon may be very natural in this context. The matter can influence the background spacetime, known as backreaction of matter. Similarly the spacetime effect can influence matter. Therefore, the thermal nature of the horizon must backreact on the matter content of the spacetime and for the thermadynamically stable configuration of horizon-matter system both the sub-systems must be at the same temperature.

\section{Black hole and thermal Wightman functions}
The static black hole metric in two dimensions can be taken in terms of Schwarzschild (static) coordinates as
\begin{equation}
ds^2 =  f(r) dt^2 - \frac{dr^2}{f(r)}~.
\label{B1}
\end{equation}
The Kruskal null-null coordinates are defined as
\begin{equation}
\bar{u} = -\frac{1}{\kappa} e^{-\kappa u}~; \bar{v} = \frac{1}{\kappa} e^{\kappa v}~,
\label{B2}    
\end{equation}
where $u = t-r^*$ and $v=t+r^*$, with $r^*$ defined by $dr^* = dr/f(r)$. Here we consider those class of black holes which are asymptotically flat, i.e. $f(r)\to 1$ for $r\to\infty$. The location of the horizon $r_H$ is determined through $f(r=r_H)=0$. $\kappa = f'(r_H)/2$ is the surface gravity. However one must note that it is possible to have more than one real solution for $r$ corresponding to equation $f(r)=0$. Among these values of $r$, the largest one determines the location of the event horizon $r_H$. Also remember that the two dimensional spacetime is conformally flat.

There are various null-null coordinates that are important for our analysis in which the metric (\ref{B1}) takes the manifestly conformally flat structure. In static null-null coordinates ({\it i.e.}, in terms of ($u,v$)), the metric takes the following form:
\begin{equation}
ds^2 =  \frac{f(r)}{2} \Big(dudv+dvdu\Big)~.
\label{B3}    
\end{equation}
In the Kruskal outgoing null coordinate $\bar{u}$ and Schwarzschild ingoing null coordinate $v$ the metric is given by
\begin{equation}
ds^2 =  -\frac{f(r)}{2\kappa\bar{u}} \Big(d\bar{u}dv+dvd\bar{u}\Big)~.
\label{B4}    
\end{equation}
Finally, in Kruskal null-null coordinates ($\bar{u},\bar{v}$) the black hole metric reduces to the following form:
\begin{equation}
ds^2 =  -\frac{f(r)}{2\kappa^2\bar{u}\bar{v}} \Big(d\bar{u}d\bar{v}+d\bar{v}d\bar{u}\Big)~.
\label{B5}    
\end{equation}
Note that the metrics (\ref{B3}), (\ref{B4}) and (\ref{B5}) are all cononformally flat, that is, these can be expressed as $g_{ab} (x) = \Omega^2 (x) \eta_{ab}$, where $\Omega^2(x)$ is the conformal factor and $\eta_{ab}$ is the Minkowski metric. Corresponding to the solutions of field equation (e.g. Klein-Gordon equation) under these forms of metric lead to various vacuum states when the field is at zero temperature.
{\it Boulware vacuum} is defined with respect to field modes in (\ref{B3}). 
The {\it Unruh vacuum} is defined with respect to modes in (\ref{B4}).
Finally, the {\it Kruskal vacuum} or {\it Hartle-Hawking vacuum} is described with respect to field modes in metric (\ref{B5}).

It is well known that in a conformally flat two dimensional spacetime, the two-point correlation functions for a massless real scalar field are given by a form which is identical to that of Minkowski spacetime (see \cite{Birrell:1982ix} for details). 
Therefore those for thermal states, defined on a confomally flat spacetime can be easily obtained by knowing the thermal correlation on Minkowski spacetime. For our spacetime (\ref{B1}), keeping the similarity to the nomenclature of the vacuua, we call the corresponding thermal states as Boulware thermal state, Unruh thermal state and Kruskal thermal state, respectively. However the relevant quantities in these thermal states boil down to their corresponding quantities in the usual vacuum states when the field temperature is considered to be vanishing one. The thermal Wightman function in a two dimensional Minkowski spacetime of the form $dS_M^2 = (1/2)(dUdV+dVdU)$, with respect to Minkowski thermal state, is obtained in \cite{Chowdhury:2019set}:
\begin{eqnarray}
&&G_{\beta}(U_2,V_2;U_1,V_1) 
\nonumber
\\
&&= -\frac{1}{4\pi} \Big( \ln\Big[1 - e^{-\frac{2\pi}{\beta} \Delta U}\Big] + \ln\Big[1 - e^{-\frac{2\pi}{\beta} \Delta V}\Big]\Big)~,
\label{B6}    
\end{eqnarray}
where $\Delta U = U_2-U_1$, $\Delta V = V_2-V_1$ and $\beta$ denotes the inverse of the temperature. As mentioned above, the required forms of Wightman functions corresponding to our thermal states will be identical in form to (\ref{B6}) with $U$ and $V$ are now replaced by null coordinates associated with respective field modes.
Therefore, the same in Boulware, Unruh and Kruskal thermal states are obtained as
\begin{eqnarray}
&&G_{\beta}^{(B)} = -\frac{1}{4\pi} \Big( \ln\Big[1 - e^{-\frac{2\pi}{\beta} \Delta u}\Big] + \ln\Big[1 - e^{-\frac{2\pi}{\beta} \Delta v}\Big]\Big)~;
\nonumber
\\
&&G_{\beta}^{(U)} = -\frac{1}{4\pi} \Big( \ln\Big[1 - e^{-\frac{2\pi}{\beta} \Delta \bar{u}}\Big] + \ln\Big[1 - e^{-\frac{2\pi}{\beta} \Delta v}\Big]\Big)~;
\nonumber
\\
&&G_{\beta}^{(K)} = -\frac{1}{4\pi} \Big( \ln\Big[1 - e^{-\frac{2\pi}{\beta} \Delta \bar{u}}\Big] + \ln\Big[1 - e^{-\frac{2\pi}{\beta} \Delta \bar{v}}\Big]\Big)~.
\nonumber
\\
\label{B7}    
\end{eqnarray}
We call these thermal Wightman functions.
In the zero temperature limit ({\it {i.e.}} $\beta\to \infty$), ignoring the well known infrared divergence in $(1+1)$-spacetime dimensions, the above ones reduce to those of non-thermal fields (see \cite{Birrell:1982ix}), which correspond to respective vacuum states.

\section{Components of renormalized energy-momentum tensor}
The general form of renormalized energy-momentum tensor, leading to trace anomaly in $(1+1)$-dimensions, is given by \cite{Birrell:1982ix}{\footnote{In \cite{Birrell:1982ix}, the last term on the right hand side of (\ref{B8}) has been taken with the negative sign. This corresponds to trace anomaly of the form $T^A_A = -\frac{^{(2)}R}{24\pi}$. However, in most of the literature the preferred form of anomaly is considered with the positive sign. Therefore to be consistent with the existing literature, for the later results we changed the sign to positive in the last term of (\ref{B8}).}}
\begin{eqnarray}
<T^A_B[g_{CD}(x)]>_{ren.} &=& \frac{\sqrt{-\eta}}{\sqrt{-g}} <T^A_B[\eta_{CD}(x)]>_{ren.} + \theta^A_B 
\nonumber
\\
&+& \frac{1}{48\pi} ^{(2)}R\delta^A_B~.
\label{B8}    
\end{eqnarray}
Here $A,B,C,D$ refers to all spacetime indices.
For a conformally flat metric of the form 
\begin{equation}
ds^2 = \frac{C_0(u_0,v_0)}{2} (du_0dv_0 + dv_0 du_0)~,
\label{B9}    
\end{equation}
in a general null-null coordinates ($u_0,v_0$), we have
\begin{eqnarray}
&&\theta_{u_0u_0} =-\frac{1}{12\pi} C_0^{1/2}\partial^2_{u_0} (C_0^{-1/2})~;
\nonumber
\\
&&\theta_{v_0v_0} =-\frac{1}{12\pi} C_0^{1/2}\partial^2_{v_0} (C_0^{-1/2})~;
\nonumber
\\
&&\theta_{u_0v_0} = 0 =\theta_{v_0u_0}~.
\label{B10}
\end{eqnarray}
Note that in the above $C_0$ has to be identified from the form of the metric (\ref{B9}) as the components of $\theta_{AB}$ are defined with respect to $(u_0,v_0)$ coordinates. Therefore when we shall calculate the components of $\theta_{AB}$ in $(u,v)$ coordinates, then $C_0$ must be identified from (\ref{B3}).
In (\ref{B8}), ~$<T^A_B[\eta_{CD}(x)]>_{ren.}$ denotes the renormalised expectation value of the stress energy tensor corresponding to the flat spacetime part of the full metric (\ref{B9}). 
The renormalized EMT (\ref{B8}) corresponds to non-vanishing of trace, but it is covariantly conserved; that is, $\nabla_A <T^A_B[g_{CD}(x)]>_{ren.} = 0$. 

In terms of static null-null components of $T^A_B$ (i.e. corresponding to coordinates considered in (\ref{B3})), the different components in Schwarzschild coordinates are given by
\begin{eqnarray}
&&T^t_t = \frac{1}{2} (T^v_u + T^u_v +  2 T^u_u) = \frac{g^{uv}}{2}(T_{uu}+ T_{vv} + 2T_{uv})~;
\nonumber
\\
&& T^r_t = T_{uu} - T_{vv}~;
\nonumber
\\
&&T^r_r = -\frac{1}{f} (T_{uu}+ T_{vv} - 2T_{uv})~,
\label{B11}
\end{eqnarray}
where for a massless real scalar field we have
\begin{eqnarray}
&&T_{uu} = \partial_u\phi \partial_u\phi~; \,\,\,\ T_{vv} = \partial_v\phi \partial_v\phi~;
\nonumber
\\
&&T_{uv} = \partial_u\phi \partial_v\phi - \frac{1}{2}g_{uv}g^{AB}\partial_A\phi\partial_B\phi~.
\label{B12}    
\end{eqnarray}
Using these, in the following we find the renormalized versions of these components in different field thermal states.

\begin{widetext}
\subsection{Unruh thermal state}
In Unruh thermal state we have
\begin{eqnarray}
&&\bra{U}T_{uu}[\eta_{AB}]\ket{U} =\lim_{u\to u'} \bra{U}\partial_u\phi (u,v) \partial_{u'}\phi(u',v')\ket{U}
= \lim_{u\to u'}\partial_u\partial_{u'} G_\beta^{(U)} (u,v;u',v') 
\nonumber
\\
&=& - \frac{\pi}{4\beta^2}\lim_{u\to u'} \frac{e^{-\kappa(u+u')}}{\sinh^2\Big(\frac{\pi\Delta\bar{u}}{\beta}\Big)}
\nonumber
\\
&=& - \frac{\pi}{4\beta^2}\lim_{u\to u'}e^{-\kappa(u+u')} \Big[\Big(\frac{\beta}{\pi\Delta\bar{u}}\Big)^2 \Big\{ 1+\frac{1}{3!}\Big(\frac{\pi\Delta\bar{u}}{\beta}\Big)^2 + \frac{1}{5!}\Big(\frac{\pi\Delta\bar{u}}{\beta}\Big)^4 + \cdots \Big\}^{-2}\Big]
\nonumber
\\
&=& - \frac{\pi}{4\beta^2}\lim_{u\to u'}e^{-\kappa(u+u')} \Big[\Big(\frac{\beta}{\pi\Delta\bar{u}}\Big)^2 - \frac{1}{3}\Big]
\nonumber
\\
&=& -\frac{\kappa^2}{16\pi}\lim_{u\to u'} \frac{1}{\sinh^2\Big(\frac{\kappa\Delta u}{2}\Big)} + \frac{\pi}{12\beta^2}e^{- 2\kappa u}
\nonumber
\\
&=& -\frac{\kappa^2}{16\pi}\lim_{u\to u'} \Big[\Big(\frac{2}{\kappa\Delta u}\Big)^2 \Big\{1+\frac{1}{3!}\Big(\frac{\kappa\Delta u}{2}\Big)^2 + \frac{1}{5!}\Big(\frac{\kappa\Delta u}{2}\Big)^4+\cdots\Big\}^{-2} \Big] 
+ \frac{\pi}{12\beta^2}e^{- 2\kappa u}
\nonumber
\\
&=& -\frac{1}{4\pi}\lim_{u\to u'} \frac{1}{(\Delta u)^2} + \frac{\kappa^2}{48\pi} + \frac{\pi}{12\beta^2}e^{- 2\kappa u}~,
\label{B13}
\end{eqnarray}
where $\Delta\bar{u} = \bar{u} - \bar{u}'$ and $\Delta u = u - u'$.
In the above the first term is divergent and this is the value for $\bra{M}T_{uu}[\eta_{AB}]\ket{M}$ when calculated with respect to Minkowski vacuum state $\ket{M}$. Therefore the renormalized quantity is evaluated as
\begin{eqnarray}
\bra{U}T_{uu}[\eta_{AB}]\ket{U}_{ren.} &=& \bra{U}T_{uu}[\eta_{AB}]\ket{U} - \bra{M}T_{uu}[\eta_{AB}]\ket{M}
\nonumber
\\
&=& \frac{\kappa^2}{48\pi} + \frac{\pi}{12\beta^2}e^{- 2\kappa u}~.
\label{B14}
\end{eqnarray}
Similarly we obtain
\begin{eqnarray}
&&\bra{U}T_{vv}[\eta_{AB}]\ket{U}_{ren.} = \frac{\pi}{12\beta^2}~;
\nonumber
\\
&& \bra{U}T_{uv}[\eta_{AB}]\ket{U}_{ren.} = 0~.
\label{B15}
\end{eqnarray}

Here we are interested to find the components of energy-momentum tensor in the static null-null coordinates (i.e. $(u,v)$) when the the Unruh thermal state is investigated. Therefore in calculating the components of $<T^A_B[g_{CD}(x)]>_{ren.}$ in ($u,v$) coordinates (chosen to express the metric in form given in (\ref{B3})), the two-point correlation function $G_{\beta}^{(U)}$ expressed in $(u,v)$ coordinates is used. So the information of the state is incorporated in the choice of thermal Wightman function. Due to this and also as was argued below Eq. (\ref{B10}), from (\ref{B3}) one identifies $C_0 =f(r)$. This we will follow in the latter calculations in connection with other thermal states as well. We will see that in the $\beta\to\infty$ limit the final results are consistent with existing ones corresponding to vacuum states. Therefore we have
\begin{eqnarray}
\theta_{uu} +\theta_{vv} &=& -\frac{1}{24\pi} C_0^{1/2} (\partial_{r^*}^2 + \partial_t^2)(C_0^{-1/2}) 
\nonumber
\\
&=& \frac{1}{48\pi} \Big[-\frac{f'^2}{2} + ff'' \Big]~.
\label{B16}    
\end{eqnarray}
Also we find $\sqrt{-\eta}=1$, $\sqrt{-g} = 1$ in Schwarzschild coordinates and $g^{uv} = 2/f(r)$ and $^{(2)}R = f''$. Here, prime denotes the derivative with respect to the radial coordinate $r$.
Therefore use of these in (\ref{B8}) yields
\begin{eqnarray}
\bra{U}T^t_t[g_{CD}]\ket{U}_{ren.} &=& \frac{\sqrt{-\eta}}{\sqrt{-g}}\bra{U}T^t_t[\eta_{CD}]\ket{U}_{ren.} + \theta_t^t + \frac{1}{48\pi} ^{(2)}R
\nonumber
\\
&=& \frac{g^{uv}}{2} \bra{U} \Big(T_{uu}[\eta_{CD}] + T_{vv}[\eta_{CD}] + 2T_{uv}[\eta_{CD}]\Big)\ket{U}_{ren.} + \frac{g^{uv}}{2}(\theta_{uu}+\theta_{vv}+2\theta_{uv})+ \frac{1}{48\pi} ^{(2)}R
\nonumber
\\
&=& \frac{1}{f(r)}\Big[\frac{\kappa^2}{48\pi} + \frac{\pi}{12\beta^2}\Big(1+e^{-2\kappa u}\Big) \Big] + \frac{1}{48\pi f(r)} \Big[-\frac{f'^2}{2} + ff'' \Big] + \frac{1}{48\pi}f''
\nonumber
\\
&=& \frac{1}{f(r)}\Big[\frac{\kappa^2}{48\pi} + \frac{\pi}{12\beta^2}\Big(1+e^{-2\kappa u}\Big) - \frac{f'^2}{96\pi} + \frac{ff''}{24\pi} \Big]~. 
\label{B17}    
\end{eqnarray}

In the same way one finds
\begin{eqnarray}
\bra{U}T^r_t[g_{CD}]\ket{U}_{ren.} &=& \frac{\sqrt{-\eta}}{\sqrt{-g}} \bra{U} \Big(T_{uu}[\eta_{CD}] - T_{vv}[\eta_{CD}]\Big)\ket{U}_{ren.} + (\theta_{uu}-\theta_{vv})
\nonumber
\\
&=& \Big[\frac{\kappa^2}{48\pi} + \frac{\pi}{12\beta^2}\Big(e^{-2\kappa u} - 1 \Big)\Big]~,
\label{BR1}    
\end{eqnarray}
and 
\begin{eqnarray}
\bra{U}T^r_r[g_{CD}]\ket{U}_{ren.} &=& -\frac{\sqrt{-\eta}}{f\sqrt{-g}} \bra{U} \Big(T_{uu}[\eta_{CD}] + T_{vv}[\eta_{CD}] - 2T_{uv}[\eta_{CD}]\Big)\ket{U}_{ren.} -\frac{1}{f}(\theta_{uu}+\theta_{vv}-2\theta_{uv}) + \frac{1}{48\pi} ^{(2)}R
\nonumber
\\
&=& -\frac{1}{f(r)}\Big[\frac{\kappa^2}{48\pi} + \frac{\pi}{12\beta^2}\Big(e^{-2\kappa u} + 1 \Big) - \frac{f'^2}{96\pi} \Big]~,
\label{BR2}    
\end{eqnarray}
For zero temperature field, these components have been calculated in \cite{Balbinot:1999vg,Das:2019aii} (correspond to Unruh vacuum), which match with the above ones for $\beta\to\infty$ limit. This manifests the consistency of our calculation.
\end{widetext}


\subsection{Boulware thermal state}
In this case, we have 
\begin{eqnarray}
&&\bra{B}T_{uu}[\eta_{AB}]\ket{B}_{ren.} 
\nonumber
\\
&=& \bra{B}T_{uu}[\eta_{AB}]\ket{B} - \bra{M}T_{uu}[\eta_{AB}]\ket{M}
\nonumber
\\
&=&  \Bigg[-\frac{1}{4\pi}\lim_{u\to u'} \frac{1}{(\Delta u)^2} + \frac{\pi}{12\beta^2}\Big]   + \frac{1}{4\pi}\lim_{u\to u'} \frac{1}{(\Delta u)^2}
\nonumber
\\
&=& \frac{\pi}{12\beta^2},
\label{B22}
\end{eqnarray}
and
\begin{eqnarray}
&&\bra{B}T_{vv}[\eta_{AB}]\ket{B}_{ren.} = \frac{\pi}{12\beta^2}~;
\nonumber
\\
&& \bra{B}T_{uv}[\eta_{AB}]\ket{B}_{ren.} = 0~.
\label{B23}
\end{eqnarray}
Therefore, we find
\begin{equation}
\bra{B}T^t_t[g_{CD}]\ket{B}_{ren.} 
= \frac{1}{f(r)}\Big[\frac{\pi}{6\beta^2} - \frac{f'^2}{96\pi} + \frac{ff''}{24\pi} \Big]~. 
\label{B24}    
\end{equation}
 The other components are evaluated as
\begin{eqnarray}
&&\bra{B}T^r_r[g_{CD}]\ket{B}_{ren.} = -\frac{1}{f(r)}\Big[\frac{\pi}{6\beta^2} - \frac{f'^2}{96\pi} \Big]~;
\nonumber
\\
&&\bra{B}T^r_t[g_{CD}]\ket{B}_{ren.} = 0~.
\label{BR3}    
\end{eqnarray}
For $\beta\to\infty$, the above components reduce to those obtained in \cite{Balbinot:1999vg}, thereby providing the consistency of our result.

\subsection{Kruskal or Hartle-Hawking thermal state}
Here we find 
\begin{eqnarray}
&&\bra{K}T_{uu}[\eta_{AB}]\ket{K}_{ren.} = \frac{\kappa^2}{48\pi} + \frac{\pi}{12\beta^2}e^{- 2\kappa u}~,
\nonumber
\\
&&\bra{K}T_{vv}[\eta_{AB}]\ket{K}_{ren.} = \frac{\kappa^2}{48\pi} + \frac{\pi}{12\beta^2}e^{ 2\kappa v}~;
\nonumber
\\
&& \bra{K}T_{uv}[\eta_{AB}]\ket{K}_{ren.} = 0~.
\label{B25}
\end{eqnarray}
Therefore one finds
\begin{eqnarray}
\bra{K}T^t_t[g_{CD}]\ket{K}_{ren.} 
&=& \frac{1}{f(r)}\Big[\frac{\kappa^2}{24\pi} + \frac{\pi}{12\beta^2}\Big(e^{-2\kappa u} + e^{2\kappa v} \Big)
\nonumber
\\
&-& \frac{f'^2}{96\pi} + \frac{ff''}{24\pi} \Big]~. 
\label{B26}    
\end{eqnarray}
The other components are
\begin{eqnarray}
&&\bra{K}T^r_r[g_{CD}]\ket{K}_{ren.} = -\frac{1}{f(r)}\Big[\frac{\kappa^2}{24\pi} + \frac{\pi}{12\beta^2}\Big(e^{-2\kappa u} + e^{2\kappa v} \Big)
\nonumber
\\
&-& \frac{f'^2}{96\pi} \Big]~;
\nonumber
\\
&&\bra{K}T^r_t[g_{CD}]\ket{K}_{ren.} = \frac{\pi}{12\beta^2} (e^{-2\kappa u} - e^{2\kappa v})~.
\label{B27}    
\end{eqnarray}
For $\beta\to\infty$, the above components again reduce to those obtained in \cite{Balbinot:1999vg}.

\section{Eneregy density and flux}
We mentioned earlier that the behavior of obtained energy-momentum components is specific to the choice of coordinates and therefore a coordinate independent description is more suitable. Hence following the previous studies \cite{Eune:2014eka,Kim:2014cja,Smerlak:2013sga,Kim:2013caa,Singh:2013pxf,Chakraborty:2015nwa} we delve onto the investigation of two scalar quantities -- the energy density and flux as measured by an observer with proper velocity $u^A$.
The energy density of the matter is given by 
\begin{equation}
\epsilon = T_A^Bu^Au_B~,
\label{E}
\end{equation}
where $u^A$ is the proper velocity of the observer that satisfies the timelike condition $u_Au^A=1$. 
The flux is given by 
\begin{equation}
F = -T_{AB}u^An^B~,
\label{F}
\end{equation}
where $n^A$ satisfies $n_An^A = -1$ and $n_Au^A=0$. For more details of the above quantities, see \cite{PhysRevD.53.1988,Ford:1993bw,PhysRevD.51.4277} where the definitions are introduced.

\subsection{Static observer}
For a static observer in Schwarzschild coordinates ($t,r$), the components of $4$-velocity is given by $u^A=(1/\sqrt{f},0)$. Therefore the energy density as measured by the static observer is given by $T^t_t$. On the other hand $n^a$ is given by $n^A = (0,-\sqrt{f})$. Therefore flux is given by $F = -\frac{1}{f}T^r_t$.

{\it Unruh thermal state: --} As energy density is determined by $T^t_t$, Eq. (\ref{B17}) gives the renormalized value of the required quantity. Note that at the horizon the above one diverges, unless the numerator 
\begin{equation}
N = \frac{\kappa^2}{48\pi} + \frac{\pi}{12\beta^2}\Big(1+e^{-2\kappa u}\Big) - \frac{f'^2}{96\pi} + \frac{ff''}{24\pi}~, 
\label{B18}
\end{equation}
vanishes at the horizon. Therefore we demand that at the horizon $N(r=r_H)=0$ for getting a finite value of energy density. Now at the horizon one has $u\to \infty, f=0$ and $f'(r=r_H) = 2\kappa$. Therefore we have
\begin{equation}
N(r=r_H) = -\frac{\kappa^2}{48\pi} + \frac{\pi}{12\beta^2}~.  
\label{B19}    
\end{equation}
Therefore vanishing of it yields the following condition
\begin{equation}
\beta = \frac{2\pi}{\kappa}~,
\label{B20}
\end{equation}
which is the inverse of Hawking temperature. In this case, the of value of $\epsilon$ at the horizon is given by
\begin{eqnarray}
\epsilon(r=r_H) &=& \lim_{r\to r_H} \frac{N'}{D'}
\nonumber
\\
&=& \lim_{r\to r_H} \Big[\frac{\pi\kappa}{6\beta^2 ff'} e^{-2\kappa u} + \frac{f''}{48\pi} + \frac{ff'''}{24\pi f'}\Big]
\nonumber
\\
&=&  \frac{f''(r_H)}{48\pi}~.
\label{B21}    
\end{eqnarray}
Note that the above value is finite and independent of the field temperature $1/\beta$. For Schwarzschild black hole $f'' = -\frac{2r_H}{r^3}$ and hence we have $\epsilon = -1/(24\pi r_H^2)$, which is negative but finite. 

For the same condition (\ref{B20}),  $\bra{U}T^r_r[g_{CD}]\ket{U}_{ren.}$ will be finite at the horizon. Consequently, its value at this location is $\frac{f''(r_H)}{48\pi}$. On the other hand,
$\bra{U}T^r_t[g_{CD}]\ket{U}_{ren.}$ is always finite and its value at the horizon is given by ($\frac{\kappa^2}{48\pi} - \frac{\pi}{12\beta^2}$). However it vanishes for the condition (\ref{B20}).
  For the choice (\ref{B20}), now one can now check that the flux in Unruh thermal state is finite at the horizon. In fact it vanishes here.

{\it Boulware thermal state: -- }
Here also for finiteness of the energy-density at horizon we must impose (\ref{B20}) and then $\epsilon(r=r_H)$ is again given by (\ref{B21}).
Like earlier $\bra{B}T^r_r[g_{CD}]\ket{B}_{ren.}$ is finite at horizon if (\ref{B20}) is satisfied and its value is then given by $\frac{f''(r_H)}{48\pi}$. Here the flux vanishes at every radial distance.

{\it Kruskal thermal state: --} 
Note that at the horizon ($u\to \infty$, $v\to -\infty$), the term within the third bracket of (\ref{B26}) vanishes and also the denominator $f(r_H)=0$. Therefore we do not need to impose any condition like (\ref{B20}). However the use of L'Hospitals rule yields
$\epsilon(r=r_H)$ as given by (\ref{B21}), which is also valid in absence of $\beta$. Hence the existence of finite temperature of the fields does not have any role in Kruskal case as long as energy is measured by a static observer stationed near the horizon.
Again $\bra{K}T^r_r[g_{CD}]\ket{K}_{ren.}$ is finite at horizon if (\ref{B20}) is satisfied and its value is then given by $\frac{f''(r_H)}{48\pi}$.
The zero temperature values can be obtained by considering the limit $\beta\to\infty$ limit, which matches the results of \cite{Balbinot:1999vg}. The flux at the horizon vanishes.

\subsection{Freely-falling observer}
The proper velocity of a freely-falling observer, moving towards the horizon, is given by\footnote{This $(1+1)$-spacetime dimensional result can be obtained from the same in $(3+1)$-dimensional case (see discussion on page number $316$ (section $7.4$) of \cite{Padmanabhan:2010zzb} for $(3+1)$-dimensional result) by imposing vanishing of angular momentum.}
\begin{equation}
u^A = \Big(\frac{E}{f(r)}, - (E^2-f(r))^{1/2} \Big)~, 
\label{B30}   
\end{equation}
where $E$ is the conserved quantity, identified as the energy per unit mass of the observer. The energy density is given by 
\begin{equation}
\epsilon = T_{AB} u^A u^B = \frac{E^2}{f}T^t_t + \frac{2E}{f^2} (E^2-f)^{1/2}T^r_t - \frac{E^2-f}{f} T^r_r~.
\label{B31}    
\end{equation}
The flux is given by (\ref{F}). For in-falling observer we find
\begin{equation}
n^A= \Big(-\frac{(E^2-f)^{1/2}}{f},E \Big)~.
\label{B32}
\end{equation}
Then the flux is calculated from
\begin{equation}
F =  \frac{E(E^2-f)^{1/2}}{f} \Big(T^t_t - T^r_r\Big) + \frac{2E^2-f}{f^2}T^r_t~.
\label{B33}    
\end{equation}

{\it Unruh thermal state: --} The energy density as measured by the in-falling observer is given by
\begin{eqnarray}
&&\epsilon^{(U)} = \bra{U}T_{AB}[g_{CD}]\ket{U} u^A u^B
\nonumber
\\
&=& \frac{E^2}{f^2} \Big[\frac{\kappa^2}{48\pi} + \frac{\pi}{12\beta^2}\Big(1+e^{-2\kappa u}\Big) - \frac{f'^2}{96\pi} + \frac{ff''}{24\pi}\Big]
\nonumber
\\
&+& \frac{2E}{f^2} (E^2 - f)^{1/2}\Big[\frac{\kappa^2}{48\pi} + \frac{\pi}{12\beta^2}\Big(e^{-2\kappa u} - 1 \Big)\Big]
\nonumber
\\
&+& \frac{E^2-f}{f^2} \Big[\frac{\kappa^2}{48\pi} + \frac{\pi}{12\beta^2}\Big(1+e^{-2\kappa u}\Big) - \frac{f'^2}{96\pi}\Big]~.
\label{B34}
\end{eqnarray}
The flux is
\begin{eqnarray}
F^{(U)} &&= \frac{E(E^2-f)^{1/2}}{f^2} \Big[\frac{\kappa^2}{24\pi} + \frac{\pi}{6\beta^2}\Big(1+e^{-2\kappa u}\Big) 
\nonumber
\\
&&- \frac{f'^2}{48\pi} + \frac{ff''}{24\pi} \Big]
\nonumber
\\
&&+ \frac{2E^2-f}{f^2}\Big[\frac{\kappa^2}{48\pi} + \frac{\pi}{12\beta^2}\Big(e^{-2\kappa u} - 1 \Big)\Big]~.
\label{B35}
\end{eqnarray}
Note that the terms in each third bracket in Eq. (\ref{B34}), at the horizon ($u\to \infty$ and $f(r_H)=0$), cancel each other for condition (\ref{B20}). Similar result is also true for the flux. Therefore, the energy density and the flux can be finite at the horizon when the matter (the scalar fields) is in thermal equilibrium with the horizon temperature.
Using L'Hospital rule one finds
\begin{eqnarray}
&&\epsilon^{(U)} = \lim_{r\to r_H}\frac{1}{2f'}\Big[\frac{E^2f'''}{24\pi} + \frac{f'f''}{48\pi}\Big] 
\nonumber
\\
&&- \lim_{r\to r_H}\frac{E}{2f}(E^2-f)^{-1/2}\Big[\frac{\kappa^2}{48\pi}\frac{\pi}{12\beta^2}(e^{-2\kappa u} - 1)\Big]
\nonumber
\\
&& - \lim_{r\to r_H} \frac{1}{2f}\Big[\frac{\kappa^2}{48\pi} + \frac{\pi}{12\beta^2}(1+e^{-2\kappa u}) - \frac{f'^2}{96\pi}\Big]
\nonumber
\\
&& = \lim_{r\to r_H}\frac{1}{2f'}\Big[\frac{E^2f'''}{24\pi} + \frac{f'f''}{48\pi}\Big] + \lim_{r\to r_H} \frac{1}{2f'} \frac{f'f''}{48\pi}
\nonumber
\\
&& = \lim_{r\to r_H} \frac{1}{2f'}\Big[\frac{E^2f'''}{24\pi} + \frac{f'f''}{24\pi}\Big]
\nonumber
\\
&& = \frac{1}{4\kappa}\Big(\frac{E^2f'''(r_H)}{24\pi}+\frac{\kappa f''(r_H)}{12\pi}\Big)~.
\label{B46}
\end{eqnarray}
Similarly, the flux at the horizon is given by
\begin{equation}
F^{(U)} = \frac{E^2f'''(r_H)}{96\pi\kappa}~.
\label{B47}
\end{equation}

{\it Boulware thermal state: --}
In this state the energy density and the flux are found as
\begin{eqnarray}
\epsilon^{(B)} &=& \frac{E^2}{f^2}\Big[\frac{\pi}{6\beta^2} - \frac{f'^2}{96\pi} + \frac{ff''}{24\pi} \Big] 
\nonumber
\\
&+& \frac{E^2-f}{f^2} \Big[\frac{\pi}{6\beta^2} - \frac{f'^2}{96\pi} \Big]~,
\label{B36}
\end{eqnarray}
and
\begin{eqnarray}
F^{(B)} &=& \frac{E(E^2-f)^{1/2}}{f^2}\Big[\frac{\pi}{3\beta^2} - \frac{f'^2}{48\pi} + \frac{ff''}{24\pi} \Big]~, 
\label{B37}
\end{eqnarray}
respectively. These are also finite at the horizon for the condition (\ref{B20}).
The values of these at the horizon are found to be (\ref{B46}) and (\ref{B47}), respectively.

{\it Hartle-Hawking thermal state: --}
In this state, the energy density and the flux are given by
\begin{eqnarray}
\epsilon^{(K)} &=& \frac{E^2}{f^2}\Big[\frac{\kappa^2}{24\pi} + \frac{\pi}{12\beta^2}\Big(e^{-2\kappa u} + e^{2\kappa v} \Big)
- \frac{f'^2}{96\pi} + \frac{ff''}{24\pi} \Big]
\nonumber
\\
&+& \frac{E^2-f}{f^2} \Big[\frac{\kappa^2}{24\pi} + \frac{\pi}{12\beta^2}\Big(e^{-2\kappa u} + e^{2\kappa v} \Big) - \frac{f'^2}{96\pi} \Big]
\nonumber
\\
&+& \frac{2E}{f^2} (E^2-f)^{1/2}\frac{\pi}{12\beta^2} (e^{-2\kappa u} - e^{2\kappa v})~;
\label{B38}
\end{eqnarray}
and 
\begin{eqnarray}
F^{(K)} &=& \frac{E(E^2-f)^{1/2}}{f^2}\Big[\frac{\kappa^2}{12\pi} + \frac{\pi}{6\beta^2}\Big(e^{-2\kappa u} + e^{2\kappa v} \Big)
\nonumber
\\
&-& \frac{f'^2}{48\pi} + \frac{ff''}{24\pi} \Big]
+ \frac{2E^2-f}{f^2}\frac{\pi}{12\beta^2} (e^{-2\kappa u} - e^{2\kappa v})~.
\nonumber
\\
\label{B39}
\end{eqnarray}
Contrary to the earlier two states, the present one leads to finite values of energy density and flux at the horizon irrespective of the temperature of the matter field. These are same as Unruh and Boulware cases and are given by (\ref{B46}) and (\ref{B47}), respectively. The results in Kruskal case are independent of field temperature for the freely-infalling observer -- was also noticed for the static observer as well. 

Note that in all the sates, the values of energy-density and flux, as seen by freely falling observer, are the same. For Schwarzschild black hole the above ones are
\begin{equation}
\epsilon = \frac{1}{8\pi r_H^2}(E^2-\frac{1}{3})~;\,\,\,\ F = \frac{E^2}{8\pi r_H^2}~. 
\label{R1}
\end{equation}
Note that the flux satisfies $F\geq 0$. 
Geodesic equation shows that $(dr/dt)^2 = (f(r)/E)^2 (E^2-f(r))$. Therefore, if the observer starts falling with a zero initial velocity from $r=r_i$, then we have $E = \sqrt{f(r_i)}$. Then the flux $F$ in (\ref{R1}) vanishes only when the observer starts from initial position which is almost on the horizon (as then $E=\sqrt{f(r_i)} = 0$). On the other hand, the energy density depending on the initial position of the infalling observer, can be both positive and negative. The maximum value of $E$ is unity when $r_i\to\infty$ and then $\epsilon = 1/(12\pi r_H^2)$. Its minimum value is $E=0$ when $r_i=r_H$ and then $\epsilon = -1/(24\pi r_H^2)$.  There is a critical value of $E$, given by $E_c = 1/\sqrt{3}$, around which the energy-density changes sign and at this critical value $\epsilon$ vanishes. The critical initial position of the observer is then given by $r^c_i = (3/2)r_H$. 

Interestingly, condition (\ref{B20}) was also reported in \cite{Milanesi:2005cs} in a completely different scenario. They observed that for the integrability of energy density at the vanishing of brick wall thickness in the ``brick wall'' approach, one needs such a condition. Furthermore, the calculation was performed by taking into account the Boulware state polarization. However, we found the generality of the condition -- it is applicable in Unruh and Boulware as well as in Hartle-Hawking states and the origin is through the finiteness of energy-density and flux as measured by two types of frame at the horizon. 

\section{Conclusion}
After constructing a matter field coupled with spacetime geometry, it is essential to construct physically meaningful observables. To do this, in literature, we find two standard approaches. As mentioned in the Introduction, here, we have not introduced particle detectors. Instead, we followed a more covariant approach in which regularized energy momentum tensor of the quantum field in the given quantum state plays the central role. The characteristic features of regularized energy momentum tensor in Boulware, Unruh and Hartle-Hawking states
for a zero temperature conformal massless scalar field propagating in the Schwarzschild spacetime has been studied even relatively recently (see \cite{Balbinot:1999vg}). In the present article, we made two fold generalizations: one by taking arbitrary radial metric function (assumed to vanish at exactly one radius) instead of Schwarzschild and another by taking thermal scalar field instead of usual zero temperature.

In our study, we took two special class of observers -- one is static and another one is freely falling towards horizon. For both the frames, the energy-density and flux at the horizon were found to be finite in Unruh and Boulware thermal state if the matter is in thermal equilibrium with the horizon, at temperature given by the Hawking expression. However, in Hartle-Hawking thermal state, those are regular at the horizon irrespective of the value of $\beta$. Particularly for the Schwarzschild black hole, we saw that the freely falling frame can have a critical initial position $r^c_i=(3/2)r_H$ where the energy-density vanishes. Around this the same is positive when $r_i>r^c_i$ and negative when $r_i<r^c_i$. In summary, for the presence of $\beta = (2\pi)/\kappa$, there is a possibility of finding finite values of energy-density and flux at the horizon. This sometimes is not possible in the limit $\beta\to\infty$; {\it{i.e.}} when the matter is at zero temperature.



We feel that the present study with finite temperature field theory shows important aspects related to role of thermal states in understanding the structure of renormalized stress-tensor near a black hole horizon. Apart from the above interesting observations, the value of the critical initial position for the freely-falling frame in Schwarzschild black hole spacetime is very interesting. This coincides with the radius of the photon sphere. But we need to be careful as the latter one is found for $(3+1)$-dimensional Schwarzschild black hole. Therefore such a resemblance can be a mare coincidence; however it needs further investigation to find any possible connection between these two completely orthogonal subjects of investigation.

\vskip 5mm
\begin{acknowledgments}
 BRM is supported by Science and Engineering Research Board (SERB), Department of Science $\&$ Technology (DST), Government of India, under the scheme Core Research Grant (File no. CRG/2020/000616). The authors thank Suprit Singh for providing useful comments.
\end{acknowledgments}

\bibliographystyle{apsrev}

\bibliography{bibtexfile}

\end{document}